\documentclass[%
 reprint,
 superscriptaddress,
 amsmath,amssymb,
aps,
pra,
]{revtex4-1}
\usepackage{graphicx}
\usepackage{subfigure}
\usepackage{amsmath}
\usepackage{amssymb}
\usepackage{tikz}
\usepackage{dcolumn}
\usepackage{bm}
\usepackage{blindtext}
\usepackage{listings}
\usetikzlibrary{quantikz}
\usepackage[colorlinks,linkcolor=blue,anchorcolor=blue,citecolor=blue]{hyperref}

\begin{document}
\title{Variational Quantum Algorithms for Steady States of Open Quantum Systems}

\author{Huan-Yu Liu}
\affiliation{Key Laboratory of Quantum Information, Chinese Academy of Sciences, School of Physics, University of Science and Technology of China, Hefei, Anhui, 230026, P. R. China}
\affiliation{CAS Center For Excellence in Quantum Information and Quantum Physics, University of Science and Technology of China, Hefei, Anhui, 230026, P. R. China}

\author{Tai-Ping Sun}
\affiliation{Key Laboratory of Quantum Information, Chinese Academy of Sciences, School of Physics, University of Science and Technology of China, Hefei, Anhui, 230026, P. R. China}
\affiliation{CAS Center For Excellence in Quantum Information and Quantum Physics, University of Science and Technology of China, Hefei, Anhui, 230026, P. R. China}

\author{Yu-Chun Wu}
\email{wuyuchun@ustc.edu.cn}
\affiliation{Key Laboratory of Quantum Information, Chinese Academy of Sciences, School of Physics, University of Science and Technology of China, Hefei, Anhui, 230026, P. R. China}
\affiliation{CAS Center For Excellence in Quantum Information and Quantum Physics, University of Science and Technology of China, Hefei, Anhui, 230026, P. R. China}

\author{Guo-Ping Guo}
\affiliation{Key Laboratory of Quantum Information, Chinese Academy of Sciences, School of Physics, University of Science and Technology of China, Hefei, Anhui, 230026, P. R. China}
\affiliation{CAS Center For Excellence in Quantum Information and Quantum Physics, University of Science and Technology of China, Hefei, Anhui, 230026, P. R. China}
\affiliation{Origin Quantum Computing Hefei, Anhui 230026, P. R. China}
\date{\today}
\begin{abstract}
Solving problems related to open quantum systems has attracted many interests. Here, we propose a variational quantum algorithm to find the steady state of open quantum systems. In this algorithm, we employ parameterized quantum circuits to prepare the purification of the steady state and define the cost function based on the Lindblad master equation, which can be efficiently evaluated with quantum circuits. Then we optimize the parameters of the quantum circuit to find the steady state. Numerical simulations are performed on the one-dimensional transverses field Ising model with dissipative channels. The result showed that the fidelity between the optimal mixed state and the true steady state is over 99\%. This algorithm is derived from the natural idea of expressing mixed states with purification and provides a reference for the study of open quantum systems.
\end{abstract}

\maketitle
\section{Introduction}

The open quantum system includes the interaction between the system and the environment. This interaction makes the state of the system have dissipation in addition to its evolution under the Hamiltonian of the system, which can be described by the Lindblad master equation \cite{lindblad}. Solving problems related to open quantum systems is of great theoretical and experimental significance \cite{hsoqs1,hsoqs2,hsoqs3,theory}, the key of which is to solve the steady state density matrix. This can be achieved by continuously iterating the master equation using classical computation. However, the dimension of the Hilbert space grows exponentially with the size of the system, making it difficult even for a medium-sized system.

The development of quantum computation gives us confidence that quantum computing can surpass its classical counterparts on certain issues. However, due to the limitation of the number of qubits and the fidelity of quantum operations, we are still far away from the age of fault-tolerant quantum computation. In the near future, we would just get access to the Noisy Intermediate-Scale
Quantum (NISQ) devices \cite{nisq1,nisq2} with dozens to hundreds of qubits available. Over the years, an NISQ-friendly algorithm, the variational quantum algorithm, has been receiving extensive attention. Usually for a task, we employ parameterized quantum circuits (PQC) to prepare ansatz, and estimates the cost function through quantum measurement with the quantum processor. Subsequently,  a classical optimizer is required to update the parameters based on the measured data to minimize the cost function. These steps are performed in a loop between the two parts until the cost function converges, returning a set of optimal parameters. Successful examples of variational quantum algorithms include variational quantum eigensolver (VQE) \cite{vqe1,vqe2,vqe3,vqe4,vqe5} and variational quantum simulation (VQS) \cite{vqs1,vqs2}, whose purposes are to find the ground state of Hamiltonian $H$ and simulate the state evolution of the systems, respectively.

In 2019, there are some works on the time evolution and steady state ansatz construction for open quantum systems based on the Restricted Boltzmann Machine (RBM) \cite{rbmr,rbmn} architecture. In those works, RBMs are used to represent the purification of the mixed state or the pure states in the ensemble \cite{work2,work3,work4}. A different idea is to represent the $n^2$  elements of the $n$-qubit density matrix with a $2n$-qubit pure state \cite{work1}: $\rho=\sum_{ij} \rho_{ij} |i\rangle \langle j| \rightarrow |\rho\rangle = \frac 1C \sum_{ij} \rho_{ij} |i\rangle \otimes |j\rangle$. Then the function of the Lindbladian $\mathcal{L}\rho$ is transformed into a "Hamiltonian": $\mathcal{L}|\rho\rangle$. And a variational quantum algorithm based on the idea was proposed in the next year \cite{dvqe}, where the Hermitian and positive semi-definite properties of the density matrix can be preserved through special design of the ansatz. However, the inconsistency of the re-normalization condition between the two types of representation encounters troubles with quantum measurement. In fact, the natural idea for this problem would be to generate the purification of the mixed state with PQC, but how to transform the Lindbladian into local operators, and therefore, how to efficiently measure the cost function with quantum hardware becomes difficult.

Here, we make a step further to propose the variational quantum algorithms for the steady states in open quantum systems, where we use PQC to generate the purification of the mixed state. The cost function is based on the Lindblad master equation and it can be decomposed into a sum of polynomial number of terms. Each of them can be evaluated using a swap test, Therefore, the cost function can be evaluated using the quantum circuits with linear scale of qubits and polynomial number of quantum gates required. We perform numerical simulations on the one-dimensional dessipative transverse field Ising model to show its reliability.

The rest of this paper is organized as follows:
In Sec. \ref{background}, we will introduce the Lindblad master equation and the steady state.
In Sec. \ref{method}, we give the method and resource estimation. We will introduce the ansatz, give the decomposition of cost function, introduce the swap test, explain the optimization sketch of parameters and estimate the quantum gate and qubit complexity.
In Sec. \ref{result}, we will give the result of the numerical simulation of the algorithm to show its correctness.
A summary and discussion will be given in Sec. \ref{conclustion}.

\section{Lindblad master equation and the steady state}\label{background}

Open quantum system considers the situation where the system interacts with its environments, in which case, the state of the system is a mixed states and can be represented with a density matrix:
\begin{equation}\label{dm}
  \rho = \sum_i p_i |\phi_i\rangle \langle \phi_i|,
\end{equation}
where $\{|\phi_i\rangle\}$ are pure states in the ensemble and $\{p_i\}$ are their corresponding occurring probabilities. The system evolves with dissipation due to the interaction, which could be described by the Lindblad master equation \cite{lindblad}:
\begin{equation}\label{s2e1}
\begin{aligned}
  \frac{d\rho}{dt} =& \mathcal{L}\rho \\
  :=& -i[H,\rho] +   \sum_i \gamma_i \left(c_i\rho c_i^+ - \frac  12 c_i^+c_i\rho-\frac 12 \rho c_i^+c_i\right).
  \end{aligned}
\end{equation}
where $\rho$ and $H$ are the system's state and Hamiltonian, respectively. $\gamma_i$ and $c_i$ are the dissipation rate and the jump operator for the i-th channel and $[H,\rho]=H\rho-\rho H$. The steady state satisfies:
\begin{equation}\label{s2e2}
  \mathcal{L}\rho_{SS} = 0,
\end{equation}
which indicates that once the state of the system is $\rho_{SS}$, then it will be invariable because of $d\rho/dt=0$. We can view the steady state as:
\begin{equation}\label{s2e10}
  \rho_{SS} = \lim_{t\to\infty} \rho(t).
\end{equation}
Therefore we can get the steady state by repeated iterations of the master equation from an initial state. However, the complexity of Hamiltonian simulation for open quantum systems makes it unpractical. Therefore, approximate method is necessary.
Here we will focus on the case where the steady state is unique. Indeed for typical systems with finite dimension of Hilbert space this is acceptable \cite{unique,unique1,unique2}.

\section{Method}\label{method}
\subsection{Ansatz}

The ansatz selection is a core part of the variational quantum algorithm. With a set of variational parameters $\{\theta\}$ in the PQC $U(\theta)$ and an initial state $|0\rangle$, generally the ansatz is: $|\psi(\theta)\rangle=U(\theta)|0\rangle$. Information-based ansatz, like unitary coupled-cluster method in quantum chemistry simulations and quantum alternating operator ansatz in combinatorial optimization problems, can take the symmetry or other properties of the system into consideration. However, these methods always cause deep circuit depth when applied on quantum circuits. While in the cases where we know little information about the system, we require that the PQC has a strong expressive power, where the solution can be found when varying the parameters, even though a cost of resource for optimization is needed.

\begin{figure}[h]
  \centering
  \begin{tikzpicture}
    \node[scale=1]{
    \begin{tikzcd}
    \lstick{$|0\rangle$} & \gate{R_z}\gategroup[4,steps=7,style={dashed,
rounded corners,fill=blue!20, inner xsep=2pt},
background]{{ repeat M layers}} & \gate{R_x} & \gate{R_z} & \ctrl{1} & \qw & \qw & \gate{R_y}&\qw \\
    \lstick{$|0\rangle$} & \gate{R_z} & \gate{R_x} & \gate{R_z} & \gate{R_y} & \ctrl{1}  & \qw & \qw & \qw\\
    \lstick{$|0\rangle$} & \gate{R_z} & \gate{R_x} & \gate{R_z} & \qw & \gate{R_y} & \ctrl{1} & \qw & \qw\\
    \lstick{$|0\rangle$} & \gate{R_z} & \gate{R_x} & \gate{R_z} & \qw & \qw & \gate{R_y} & \ctrl{-3} & \qw
    \end{tikzcd}
    };
  \end{tikzpicture}
  \caption{The illustration of the hardware-efficient ansatz used in this work with an example of $2n=4$. In the circuit, each layer consists of single-qubit rotation gates and two-qubit entangling gates. And the structure repeated $M$ times to construct $U(\theta)$ to which the parameter would be different for a specific gate in each layer. The ansatz is obtained by tracing out the bottom $n=2$ qubits.}\label{mpqc}
\end{figure}
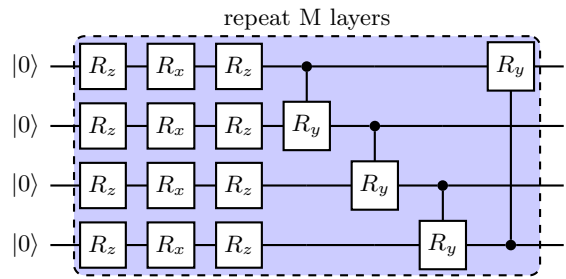

Here, we use the hardware-efficient ansatz \cite{vqe1,mpqc1,mpqc2}, which combines easily implementable single- and double-qubit operations as one layer. And the layers are repeated to increase the expressive power. Here, the structure we choose to construct the ansatz is:
\begin{equation}\label{s4e2}
  U(\theta) = \prod_{i=1}^{M} \prod_{j=1}^{N} U_{ent}(\theta_i)R_z(\theta_{ij1})R_x(\theta_{ij2})R_z(\theta_{ij3}),
\end{equation}
where $N$ and $M$ are the number of qubits and layers in the circuit, respectively. For the entangling part $U_{ent}$ we choose sequential $control-R_y$ operations. The illustration of the PQC we use is shown in Fig. \ref{mpqc} with a systems of 2 qubits as an example.

Considering the $n$-qubit mixed state formulized as Eq. (\ref{dm}), we can view the system and the environment as a closed system, where the system-environment joint state is the purification of the system's state:
\begin{equation}\label{purif}
  |\Phi\rangle = \sum_i \sqrt{p_i} |\phi_i\rangle \otimes |e_i\rangle,
\end{equation}
such that we can obtain the state of the system by tracing out the environment as:
\begin{equation*}
  Tr_E(|\Phi\rangle \langle\Phi| ) = \rho.
\end{equation*}

Note that for an $n$-qubit mixed state, we can always use another $n$ auxiliary qubits to construct its purification \cite{book}. Thus, we can first construct a $2n$-qubit pure state:
\begin{equation}\label{purification}
  |\Psi(\theta) \rangle= U(\theta)|0\rangle^{\otimes 2n},
\end{equation}
and then trace out the n auxiliary qubits to obtain the ansatz $\rho(\theta)$.

\subsection{Cost function}
After defining the ansatz of target density matrix, here we introduce the cost function $\mathcal{C}(\theta)$. The steady state satisfies Eq. (\ref{s2e2}), which indicates that in the Linbblad form for a density matrix $\rho=\sum_{ij}\rho_{ij}|i\rangle\langle j|$, we have:
\begin{equation}\label{s4e3}
  \frac{d}{dt}\rho =\mathcal{L}\rho = \sum_{ij} \frac{d}{dt} \rho_{ij} |i\rangle\langle j|.
\end{equation}
This shows that if $\mathcal{L}\rho=0$, we have $\frac{d}{dt}\rho_{ij}=0$ for all indices $i$ and $j$, such that the state evolution under the Lindblad master equation will be invariable. Therefore we define the cost function based on the Frobenius  norm of $\mathcal{L}\rho$:
\begin{equation}\label{s4e3}
  \mathcal{C}(\theta) = ||\mathcal{L}\rho(\theta)||_F^2 = \sum_{ij} |(\mathcal{L}\rho(\theta))_{ij}|^2.
\end{equation}
It is obvious that $\mathcal{C}(\theta)\geq 0$ for all $\theta$ and $\mathcal{C}(\theta)=0\Leftrightarrow \rho=\rho_{SS}$.
We observe that the Frobenius norm of a matrix can be transformed as:
\begin{equation*}
  \mathcal{C} (\theta) = || \mathcal{L} \rho ||_F^2 = tr[ (\mathcal{L} \rho)^{\dagger}\mathcal{L} \rho  ].
\end{equation*}

According to Eq. (\ref{s2e1}), where the function of Lindbladian can be viewed as a sum of terms, each of which has a unitary operator multiplies on the left and/or right of the density matrix. We can re-write the function of the Lindbladian as:
\begin{align*}
  \mathcal{L} \rho &= \sum_i f_i U_i \rho V_i \\
   (\mathcal{L} \rho)^{\dagger} &= \sum_i f_i^* V_i^{\dagger} \rho U_i^{\dagger},
\end{align*}
whre $f_i$ are complex numbers and $U_i/V_i$ are unitary operators. Then the cost function can be decomposed as:
\begin{equation*}
tr[ (\mathcal{L} \rho)^{\dagger}\mathcal{L} \rho  ] = tr( \sum_{ij} f_i^*f_j V_i^{\dagger} \rho U_i^{\dagger} U_j\rho V_j  ).
\end{equation*}
Since $tr(AB)=tr(BA)$, finally the cost function is transformed to the following form:
\begin{equation}\label{costfinal}
  \mathcal{C} (\theta) = \sum_{ij}  f_i^*f_j tr(   \rho U_i^{\dagger} U_j \rho V_j V_i^{\dagger}      ).
\end{equation}

\subsection{Swap test}

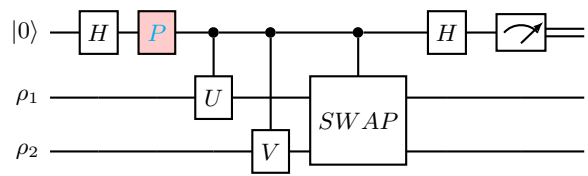
\begin{figure}[h]
    \center
    \begin{tikzpicture}
      \node[scale=1]{
      \begin{tikzcd}
      \lstick{$|0\rangle$}     &\gate{H}  &\gate[style={fill=red!20},label style=cyan]{P} &\ctrl{1} &\ctrl{2} &\ctrl{1}              &\gate{H}  &\meter{} &\cw     \\
      \lstick{$\rho_1$}  &\qw  &\qw    &\gate{U} &\qw      &\gate[wires=2]{SWAP}  &\qw       &\qw      &\qw    \\
      \lstick{$\rho_2$}  &\qw  &\qw    &\qw      &\gate{V} &                      &\qw       &\qw      &\qw
      \end{tikzcd}
      };
    \end{tikzpicture}
    \caption{The quantum circuit of the swap test to measure terms in the form of $Tr( \rho U\rho V )$. Desired values are associated with the probability of obtaining $|0\rangle$ when measuring the auxiliary qubit at the end of the circuit. When evaluating the real part, the red phase gate $P$ is not applied and we have:  $\Re[    tr(   \rho_1V\rho_2U     ) ] = 2p(0) - 1$. While for the imaginary part, the gate $P$ is applied with the relation changed: $ \Im[    tr(   \rho_1V\rho_2U     ) ] = 1-2p(0) $.   }\label{modifyswap}
\end{figure}

In Eq. (\ref{costfinal}), every term has the form of $tr( \rho U\rho V )$, which can be estimated using the swap test. The swap test can be applied to obtain the value of inner product terms, which is frequently used to evaluate functions like quantum state fidelity. The quantum circuit of the swap test to evaluate the cost function for this work is shown in Fig. \ref{modifyswap}. The auxiliary qubit is measured at the end of the circuit and we have the following relation:
\begin{equation}\label{modify}
\Re[    tr(   \rho_1V\rho_2U     ) ] = 2p(0) - 1.
\end{equation}
where $\Re[.]$  denotes the real part of a number. When we want to obtain the imaginary part we can add a phase gate $P=\begin{pmatrix}
                                               1 & 0 \\
                                               0 & i
                                             \end{pmatrix}$ after the first $H$ gate, then the relations will be:
\begin{equation*}
  \Im[    tr(   \rho_1V\rho_2U     ) ] = 1-2p(0)
\end{equation*}

\subsection{Parameter optimization}
Optimizing the parameters to minimize the cost function is quite difficult in variational quantum algorithms. For randomly selected PQC structure and initilized parameters in large quantum systems, the Barren plateau phenomenon has been studied \cite{bp1,bp2}. And it was also shown in \cite{bplocal} that the phenomenon is dependent on the locality of the cost function.  Besides, finite sampling times and quantum gate errors makes the value of measured cost fuction stochastic, which all bring troubles with the optimization process. In our numerical simulation, we use the gradient-free method, Nelder-Mead (NM) algorithm \cite{nm}, which has been used in some variational quantum algorithms \cite{vqe2,nmvqe}.  In our simulation, cost functions are evaluated using "statevector simulators" in Qiskit \cite{qiskit}, where no sampling error or gate error is considered. Several optimization methods for the stochastic optimization problem have been proposed, Then further studies will be to use targeted methods to solve the stochastic optimization problem.

The NM method starts with a series of randomly initial points, then obtains a new point to replace the worst point. These steps stop when the change tolerance is satisfied or the maximum iteration times is achieved. The method written in Scipy \cite{scipy} has a default value of the maximum iteration steps: $200\times d$, where d is the number of parameters. We find that instead of changing this default number, repeating this algorithm for several times obtains a better result. Therefore, in our test, we repeat the algorithms for several times to achieve sufficient optimization.

\subsection{Resource estimation}
\paragraph{Qubits}
For an $n$ qubit system, we use $2n$ qubits to construct the purification. In the swap test, we need two copies and one auxiliary qubit. The total number of qubits required to evaluate the cost function is $4n+1$. Single-qubit measurement is enough in the swap test.

\paragraph{Number of terms in the cost function}
Beside the qubits, the number of terms for measuring is also important for this algorithm, which greatly influences the cost function evaluation efficiency and sampling accuracy. Suppose the Hamiltonian and $m$ dissipative  operators can be expressed as a linear combination of $q_h$ and $q_l$ easily implementable unitary operators ($q_h$ and $q_l$ are assumed to belong to $O(poly(n))$ naturally, The totally number of terms is then about:
\begin{equation*}
  O(  (  2q_h +3mq_l  )^2   ).
\end{equation*}
Here we consider that $m\in O(poly(n))$ and then the total number is also $O(poly(n))$, which can be efficiently measured.

\paragraph{Gates and parameters}

In the hardware efficient ansatz with an $n$-qubit system and $M$-layers, there are totally $8nM$ parameters. And since each $control-R_y$ gate can be decomposed into 2 CNOT gates and 2 $R_y$ gates. Therefore, totally $5nM$ single-qubit rotation gates and $2nM$ CNOT gates are required.

In the swap test, the $O(n)$ $control-U/V$ gates can be decomposed into single- and double-qubit gates in the same order. While the $control-SWAP$ gate can be realized with $n$ $C-SWAP$ gates, where each one is equal to 3 Toffoli gates. Therefore, the total number of quantum gates needed scales linearly with the size of the system.

All in all, we can see that the resource to measure the cost function is acceptable.

\begin{figure}[h]
  \centering
  \subfigure[loss function]{\includegraphics[scale=0.5]{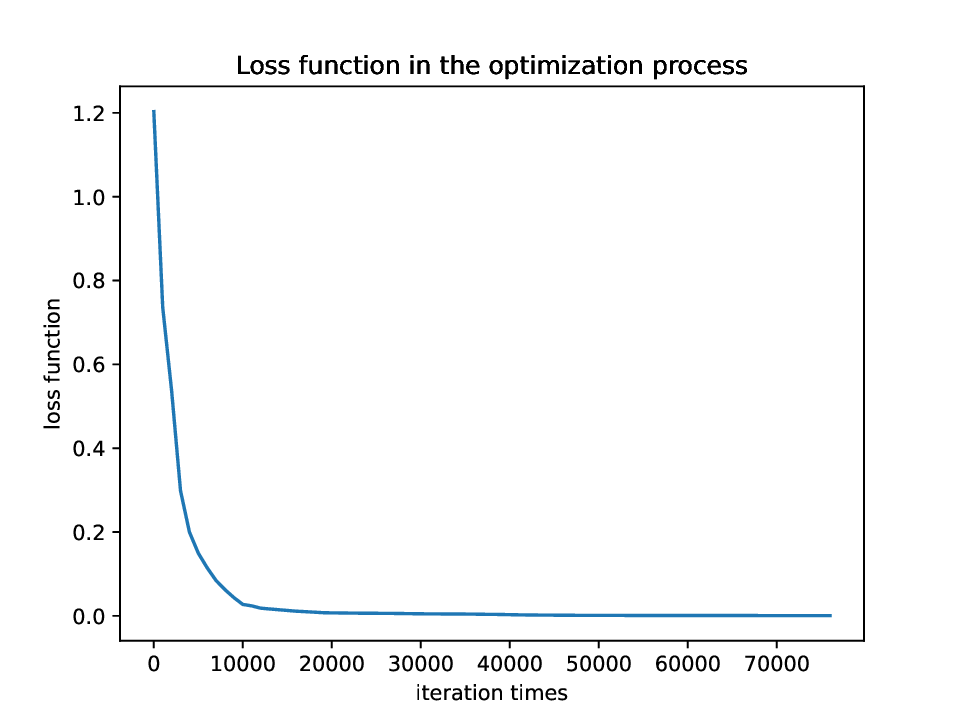}}
  \quad
  \subfigure[fidelity]{\includegraphics[scale=0.5]{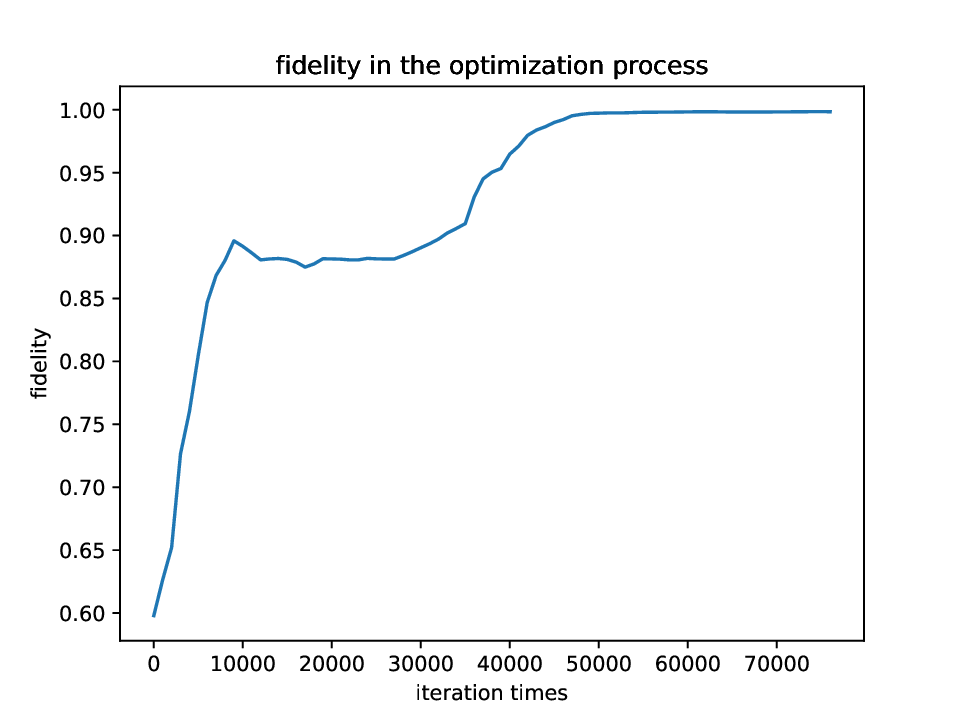}}
  \caption{Data in the optimization process for the dissipative one dimensional transverse field Ising model. (a), The values of loss function with respect to the iteration times. The total iteration times is about $8\times 10^4$ and the optimal loss function is about $1.8\times 10^{-3}$. (b): The values of fidelity between the ansatz and the stationary state density matrix obtained with QuTiP in the optimization process. The fidelity increases along with the cost function decreasing. And the optimal fidelity is over 99.8\%, which can show the validity of our method.}
  \label{lfi}
\end{figure}

\begin{figure*}[h]
\subfigure[Ising model, QuTiP]{
\includegraphics[scale=0.5]{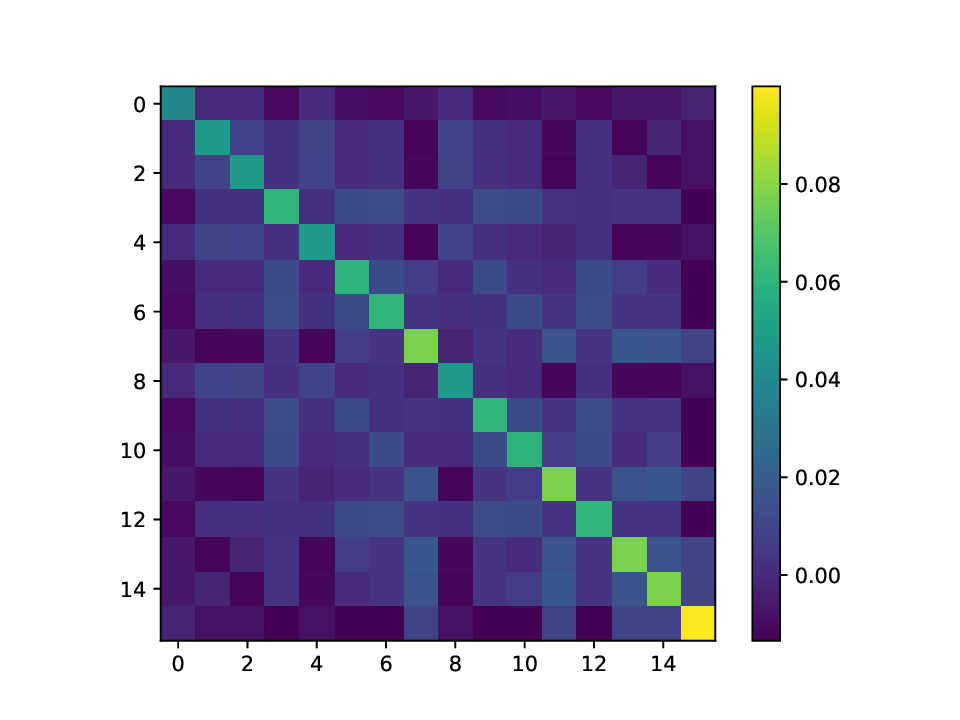}
\quad
\includegraphics[scale=0.5]{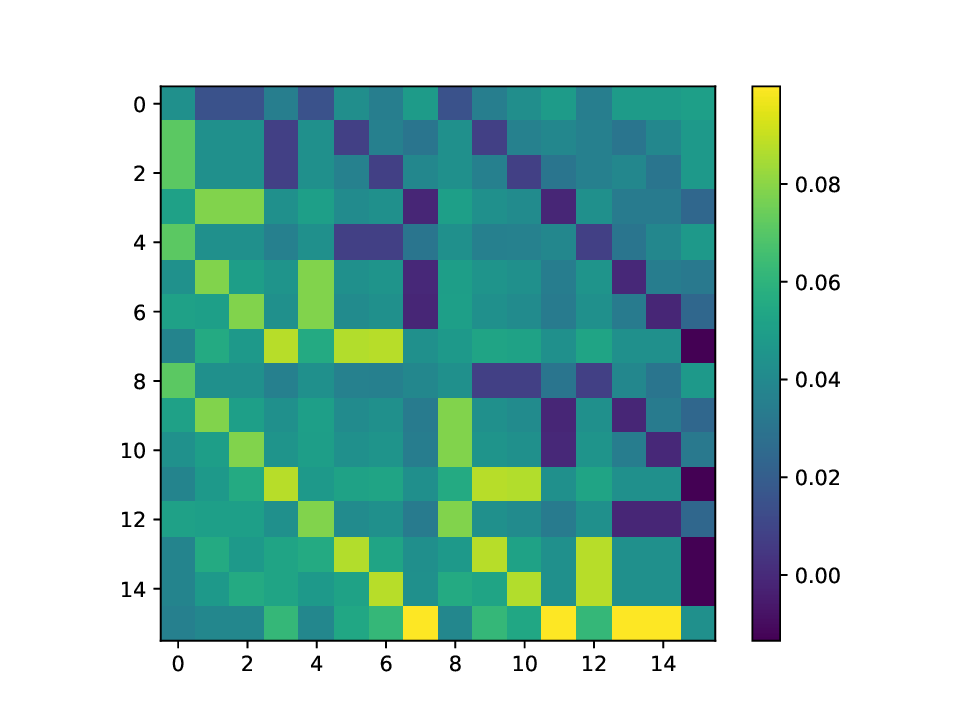}
}
\quad
\subfigure[Ising model, our method]{
\includegraphics[scale=0.5]{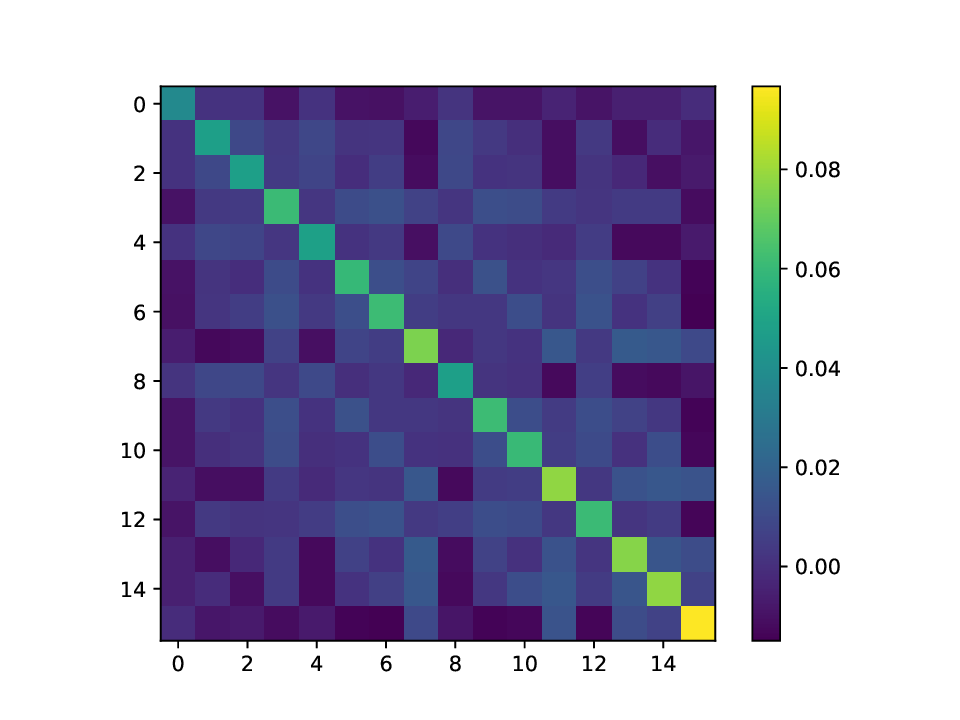}
\quad
\includegraphics[scale=0.5]{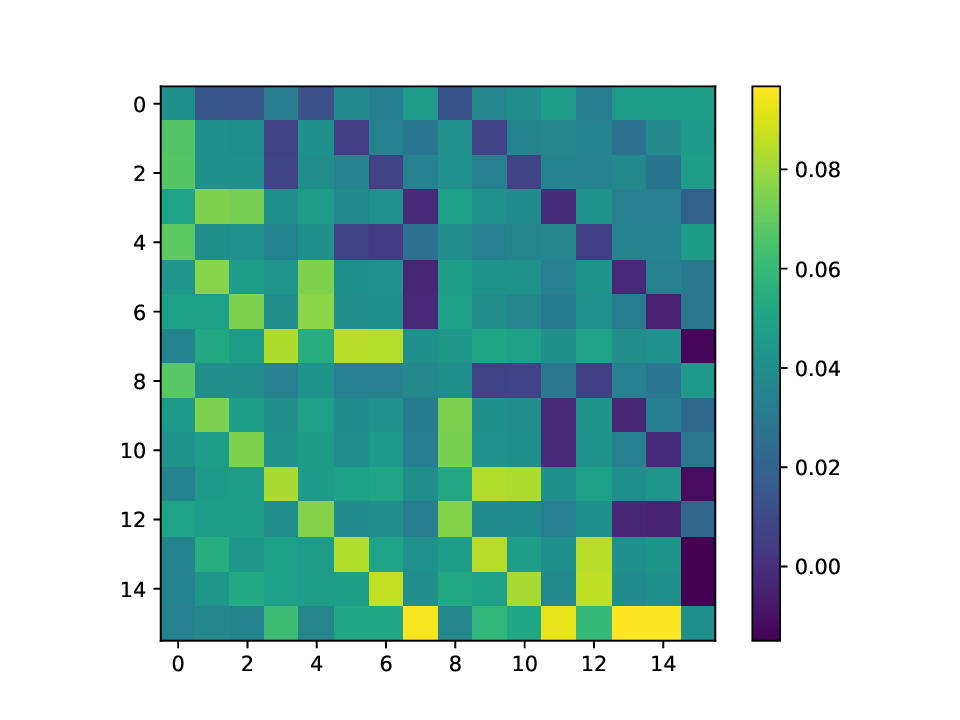}
}
\caption{The image pictures of the stationary state density matrices obtained with QuTiP and out optimal ansatz. The left and the right part represents the real and imaginary part of the density matrix, respectively. There is no visible difference, which can also show the correctness of our method.}
\label{data}
\end{figure*}

\section{Result}\label{result}
In this part we perform numerical simulations on the one dimensional dissipative transverse field Ising model. The code is written is Python. We use Qiskit and Scipy packages to process the data and optimize the parameters. And the computational package for dynamics of open quantum systems, QuTiP \cite{qutip}, to numerically obtain the steady state as a benchmark.  We use the fidelity between our optimal ansatz and the steady state obtained with QuTiP to compare their difference. The fidelity of  two density operators $\sigma$ and $\rho$ is given by:
\begin{equation}\label{s5e3}
    F(\sigma,\rho) = \left( Tr \sqrt{   \sqrt{\sigma}   \rho \sqrt{\sigma}     }   \right)^2.
\end{equation}

The Hamiltonian of the model is:
\begin{equation}\label{s5e1}
  H =\frac V4 \sum_{i=0}^{L-1}  \sigma_i^z \sigma_{i+1}^z + \frac g2 \sum_{i=0}^{L-1}\sigma_i^x,
\end{equation}
where $L$ is the length of the spin chain with $L+1$ spin particles, labeled as $0,1.\cdots,L$, $\sigma^a = \frac 12 S^a$ with $a=\{x,y,z\}$ is the Pauli operator, V and g models the nearest-neighbor interaction strength in z-axis and the amplitude of transverse field along the x-axis. The periodic boundary condition is applied, which indicates that $\sigma_L=\sigma_0$. We assume that there is only one quantum channel on every qubit. The jump operator and dissipation rate are defined as follows:
\begin{equation}\label{s5e2}
  c_i = \sigma_i^- = \frac 12 \left(\sigma_i^x - i\sigma_i^y\right), \qquad \gamma_i = \gamma.
\end{equation}

We performed our test of this model with the model parameters as $L=4$, $V=0.3$, $g=1$ and $\gamma=0.5$, respectively. We use 4 auxiliary qubits according to former analysis. The number of layers in the hardware-efficient ansatz is 4. The data in the optimization process is shown in Fig.\ref{lfi}. In Fig.\ref{lfi}(a), we plot the change of loss function with respect to the iteration times. The total iteration times is about $8\times 10^4$ and we reach the optimal loss function at about $1\times 10^{-3}$. We plot the change of fidelity of our ansatz density matrix with the one obtained with QuTiP in Fig. \ref{lfi}(b). Obviously we can see that the fidelity increases in the optimization process and the optimal fidelity is over 99.8\%. In Fig. \ref{data}, we plot the image pictures of the steady state density matrix obtained with our method and the one from QuTiP. There is no visible difference, which can show the correctness of our method.

\section{Conclusion and discussion}\label{conclustion}
In this work we have proposed a variational quantum algorithm to find the steady state of open quantum systems. We use hardware-efficient ansatz to generate the purification of the ansatz for the mixed state. The cost function is based on the Lindbladian and is decomposed into a sum of polynomial number of terms, where each of them can be evaluated using a swap test with the quantum hardware. We performed numerical simulations on the one-dimensional dissipative transverse field Ising model. The optimal fidelity between the ansatz and the one obtained with QuTiP is over 99\%, showing the reliability of our method.

Just like many other variational quantum algorithms, the limited sampling times and the noise of quantum hardware have caused a lot of troubles in the optimization process. Finding an efficient optimization method will be a possible work for our algorithm in the future. Due to the importance of solving steady states in open quantum systems and the preventive use of this type of loss function in various tasks, more applications would be found with this algorithm.

\acknowledgments
We thank Prof. Yong-Jian Han for helpful discussion.
The numerical calculations in this paper have been done on the supercomputing system in the Supercomputing Center of University of Science and Technology of Chin.
This work was supported by the National Key Research and Development Program of China (Grant No. 2016YFA0301700) and the Anhui Initiative in Quantum Information Technologies (Grant No. AHY080000).

\bibliography{ref}

\begin{thebibliography}{10}

\bibitem{lindblad}
Goran Lindblad.
\newblock On the generators of quantum dynamical semigroups.
\newblock {\em Communications in Mathematical Physics}, 48(2):119--130, 1976.

\bibitem{hsoqs1}
Julio~T Barreiro, Markus M{\"u}ller, Philipp Schindler, Daniel Nigg, Thomas
  Monz, Michael Chwalla, Markus Hennrich, Christian~F Roos, Peter Zoller, and
  Rainer Blatt.
\newblock An open-system quantum simulator with trapped ions.
\newblock {\em Nature}, 470(7335):486, 2011.

\bibitem{hsoqs2}
Ryan Sweke, Ilya Sinayskiy, and Francesco Petruccione.
\newblock Simulation of single-qubit open quantum systems.
\newblock {\em Phys. Rev. A}, 90:022331, Aug 2014.

\bibitem{hsoqs3}
Mattias Fitzpatrick, Neereja~M. Sundaresan, Andy C.~Y. Li, Jens Koch, and
  Andrew~A. Houck.
\newblock Observation of a dissipative phase transition in a one-dimensional
  circuit qed lattice.
\newblock {\em Phys. Rev. X}, 7:011016, Feb 2017.

\bibitem{theory}
Jamir Marino and Sebastian Diehl.
\newblock Quantum dynamical field theory for nonequilibrium phase transitions
  in driven open systems.
\newblock {\em Phys. Rev. B}, 94:085150, Aug 2016.

\bibitem{nisq1}
John Preskill.
\newblock Quantum computing in the nisq era and beyond.
\newblock {\em Quantum}, 2:79, 2018.

\bibitem{nisq2}
Jay Gambetta.
\newblock Benchmarking nisq-era quantum processors.
\newblock In {\em APS Meeting Abstracts}, 2019.

\bibitem{vqe1}
Abhinav Kandala, Antonio Mezzacapo, Kristan Temme, Maika Takita, Markus Brink,
  Jerry~M Chow, and Jay~M Gambetta.
\newblock Hardware-efficient variational quantum eigensolver for small
  molecules and quantum magnets.
\newblock {\em Nature}, 549(7671):242, 2017.

\bibitem{vqe2}
Alberto Peruzzo, Jarrod McClean, Peter Shadbolt, Man-Hong Yung, Xiao-Qi Zhou,
  Peter~J Love, Al{\'a}n Aspuru-Guzik, and Jeremy~L O’brien.
\newblock A variational eigenvalue solver on a photonic quantum processor.
\newblock {\em Nature communications}, 5:4213, 2014.

\bibitem{vqe3}
P.~J.~J. O'Malley, R.~Babbush, I.~D. Kivlichan, J.~Romero, J.~R. McClean,
  R.~Barends, J.~Kelly, P.~Roushan, A.~Tranter, N.~Ding, B.~Campbell, Y.~Chen,
  Z.~Chen, B.~Chiaro, A.~Dunsworth, A.~G. Fowler, E.~Jeffrey, E.~Lucero,
  A.~Megrant, J.~Y. Mutus, M.~Neeley, C.~Neill, C.~Quintana, D.~Sank,
  A.~Vainsencher, J.~Wenner, T.~C. White, P.~V. Coveney, P.~J. Love, H.~Neven,
  A.~Aspuru-Guzik, and J.~M. Martinis.
\newblock Scalable quantum simulation of molecular energies.
\newblock {\em Phys. Rev. X}, 6:031007, Jul 2016.

\bibitem{vqe4}
E.~F. Dumitrescu, A.~J. McCaskey, G.~Hagen, G.~R. Jansen, T.~D. Morris,
  T.~Papenbrock, R.~C. Pooser, D.~J. Dean, and P.~Lougovski.
\newblock Cloud quantum computing of an atomic nucleus.
\newblock {\em Phys. Rev. Lett.}, 120:210501, May 2018.

\bibitem{vqe5}
Raffaele Santagati, Jianwei Wang, Antonio~A. Gentile, Stefano Paesani, Nathan
  Wiebe, Jarrod~R. Mcclean, Sam Morley-Short, Peter~J. Shadbolt, Damien
  Bonneau, and Joshua~W. Silverstone.
\newblock Witnessing eigenstates for quantum simulation of hamiltonian spectra.
\newblock {\em Science Advances}, 4(1):eaap9646, 2018.

\bibitem{vqs1}
Xiao Yuan, Suguru Endo, Qi~Zhao, Ying Li, and Simon~C Benjamin.
\newblock Theory of variational quantum simulation.
\newblock {\em Quantum}, 3:191, 2019.

\bibitem{vqs2}
Sam McArdle, Tyson Jones, Suguru Endo, Ying Li, Simon Benjamin, and Xiao Yuan.
\newblock Variational quantum simulation of imaginary time evolution.
\newblock {\em arXiv preprint arXiv:1804.03023}, 2018.

\bibitem{rbmr}
Giuseppe Carleo, Yusuke Nomura, and Masatoshi Imada.
\newblock Constructing exact representations of quantum many-body systems with
  deep neural networks.
\newblock {\em Nature communications}, 9(1):5322, 2018.

\bibitem{rbmn}
Giacomo Torlai and Roger~G. Melko.
\newblock Latent space purification via neural density operators.
\newblock {\em Phys. Rev. Lett.}, 120:240503, Jun 2018.

\bibitem{work2}
Alexandra Nagy and Vincenzo Savona.
\newblock Variational quantum monte carlo method with a neural-network ansatz
  for open quantum systems.
\newblock {\em Phys. Rev. Lett.}, 122:250501, Jun 2019.

\bibitem{work3}
Michael~J. Hartmann and Giuseppe Carleo.
\newblock Neural-network approach to dissipative quantum many-body dynamics.
\newblock {\em Phys. Rev. Lett.}, 122:250502, Jun 2019.

\bibitem{work4}
Filippo Vicentini, Alberto Biella, Nicolas Regnault, and Cristiano Ciuti.
\newblock Variational neural-network ansatz for steady states in open quantum
  systems.
\newblock {\em Phys. Rev. Lett.}, 122:250503, Jun 2019.

\bibitem{work1}
Nobuyuki Yoshioka and Ryusuke Hamazaki.
\newblock Constructing neural stationary states for open quantum many-body
  systems.
\newblock {\em Phys. Rev. B}, 99:214306, Jun 2019.

\bibitem{dvqe}
Nobuyuki Yoshioka, Yuya~O. Nakagawa, Kosuke Mitarai, and Keisuke Fujii.
\newblock Variational quantum algorithm for nonequilibrium steady states.
\newblock {\em Phys. Rev. Research}, 2:043289, Nov 2020.

\bibitem{unique}
Davide Nigro.
\newblock On the uniqueness of the steady-state solution of the
  lindblad--gorini--kossakowski--sudarshan equation.
\newblock {\em Journal of Statistical Mechanics: Theory and Experiment},
  2019(4):043202, 2019.

\bibitem{unique1}
S.~G. Schirmer and Xiaoting Wang.
\newblock Stabilizing open quantum systems by markovian reservoir engineering.
\newblock {\em Phys. Rev. A}, 81:062306, Jun 2010.

\bibitem{unique2}
J-T Hsiang and BL~Hu.
\newblock Nonequilibrium steady state in open quantum systems: influence
  action, stochastic equation and power balance.
\newblock {\em Annals of Physics}, 362:139--169, 2015.

\bibitem{mpqc1}
Sukin Sim, Peter~D Johnson, and Alan Aspuru-Guzik.
\newblock Expressibility and entangling capability of parameterized quantum
  circuits for hybrid quantum-classical algorithms.
\newblock {\em arXiv preprint arXiv:1905.10876}, 2019.

\bibitem{mpqc2}
Yuxuan Du, Min-Hsiu Hsieh, Tongliang Liu, and Dacheng Tao.
\newblock The expressive power of parameterized quantum circuits.
\newblock {\em arXiv preprint arXiv:1810.11922}, 2018.

\bibitem{book}
Michael~A Nielsen and Isaac Chuang.
\newblock Quantum computation and quantum information, 2002.

\bibitem{bp1}
Jarrod~R McClean, Sergio Boixo, Vadim~N Smelyanskiy, Ryan Babbush, and Hartmut
  Neven.
\newblock Barren plateaus in quantum neural network training landscapes.
\newblock {\em Nature communications}, 9(1):1--6, 2018.

\bibitem{bp2}
Edward Grant, Leonard Wossnig, Mateusz Ostaszewski, and Marcello Benedetti.
\newblock An initialization strategy for addressing barren plateaus in
  parametrized quantum circuits.
\newblock {\em Quantum}, 3:214, 2019.

\bibitem{bplocal}
M.~Cerezo, Akira Sone, Tyler Volkoff, Lukasz Cincio, and Patrick~J. Coles.
\newblock Cost function dependent barren plateaus in shallow parametrized
  quantum circuits.
\newblock {\em Nature Communications}, 12(1), Mar 2021.

\bibitem{nm}
Sa{\v{s}}a Singer and John Nelder.
\newblock Nelder-mead algorithm.
\newblock {\em Scholarpedia}, 4(7):2928, 2009.

\bibitem{nmvqe}
Cornelius Hempel, Christine Maier, Jonathan Romero, Jarrod McClean, Thomas
  Monz, Heng Shen, Petar Jurcevic, Ben~P. Lanyon, Peter Love, Ryan Babbush,
  Al\'an Aspuru-Guzik, Rainer Blatt, and Christian~F. Roos.
\newblock Quantum chemistry calculations on a trapped-ion quantum simulator.
\newblock {\em Phys. Rev. X}, 8:031022, Jul 2018.

\bibitem{qiskit}
Gadi Aleksandrowicz, Thomas Alexander, Panagiotis Barkoutsos, Luciano Bello,
  Yael Ben-Haim, D~Bucher, FJ~Cabrera-Hern{\'a}ndez, J~Carballo-Franquis,
  A~Chen, CF~Chen, et~al.
\newblock Qiskit: An open-source framework for quantum computing.
\newblock {\em Accessed on: Mar}, 16, 2019.

\bibitem{scipy}
Eli Bressert.
\newblock {\em SciPy and NumPy: an overview for developers}.
\newblock " O'Reilly Media, Inc.", 2012.

\bibitem{qutip}
J~Robert Johansson, Paul~D Nation, and Franco Nori.
\newblock Qutip 2: A python framework for the dynamics of open quantum systems.
\newblock {\em Computer Physics Communications}, 184(4):1234--1240, 2013.

\end{thebibliography}
\bibliographystyle{unsrt}

\end{document}